\documentclass[useAMS,usenatbib]{mn2e}

\usepackage{graphicx}
\usepackage{epstopdf}

\title[On the nature of cosmological time]{On the nature of cosmological time}
\author[P. Magain and C. Hauret]{P. Magain$^{1}$\thanks{E-mail:
pierre.magain@ulg.ac.be } and C. Hauret$^{1}$\\
$^{1}$Institut d'Astrophysique et de G\'eophysique, Universit\'e de Liege, All\'ee du 6 Ao\^ut 19c, B-4000 Liege, Belgium}
\begin{document}

\date{}

\pagerange{\pageref{firstpage}--\pageref{lastpage}} \pubyear{2002}

\maketitle

\label{firstpage}

\begin{abstract}
Time is a parameter playing a central role in our most fundamental modeling of natural laws.  Relativity theory shows that the comparison of times measured by different clocks depends on their relative motions and on the strength of the gravitational field in which they are embedded.  In standard cosmology, the time parameter is the one measured by fundamental clocks, i.e. clocks at rest with respect to the expanding space.  This proper time is assumed to flow at a constant rate throughout the whole history of the Universe.  We make the alternative hypothesis that the rate at which cosmological time flows depends on the dynamical state of the Universe.  In thermodynamics, the arrow of time is strongly related to the second law, which states that the entropy of an isolated system will always increase with time or, at best, stay constant.  Hence, we assume that time measured by fundamental clocks is proportional to the entropy 
of the region of the Universe that is causally connected to them.  Under that simple assumption, we build a cosmological model that explains the Type Ia Supernovae data (the best cosmological standard candles) without the need for exotic dark matter nor dark energy.
\end{abstract}


\section{Introduction}

Since its introduction nearly a century ago \citep{einstein1916}, general relativity (GR) has been brilliantly confirmed by a number of observations, most notably the perihelion precession of Mercury, the gravitational redshift and the deflection of light by massive bodies.  GR has also been used in cosmology to describe the evolution of the Universe as a whole, and the model that currently gives the best description of the large-scale structure and evolution of the Universe, namely the Lambda Cold Dark Matter ($\Lambda$CDM) model, is based on the equations of GR.  However, in order to provide an accurate description of large-scale and long-term phenomena, a number of ingredients had to be added to the theory.  The motion of stars in the outskirts of spiral galaxies and the velocity dispersions of galaxies in clusters required the addition of dark matter.  The acceleration of the expansion of the Universe called for the introduction of a repulsive component called dark energy.

One may note that all the successful tests of GR are dealing with small scales and are quasi-instantaneous in terms of cosmological time.  Conversely, additional ingredients (dark matter and dark energy) are necessary to bring the models in agreement with observations dealing with very large scales and very long term phenomena.  We are thus faced with the following alternative: either these dark components are actual constituents of the Universe (in which case they should sooner or later be identified) or there is something wrong with the application of GR on very large scales and, especially, on the Universe as a whole.

The vast majority of specialists are in favour of the existence of dark matter and dark energy.  However, despite considerable effort, these dark components have not been identified yet.  Some alternative theories have been proposed, the most popular being the so-called Modified Newtonian Dynamics (MOND, \citealt{milgrom1983}).

In this paper, we want to explore another hypothesis, namely that the problems encountered when dealing with large scales are related to our fundamental understanding of time.  Special relativity has shown that the time measured by different clocks depends on their relative motions.  GR has shown that the time measured by a clock also depends on the strength of the gravitational field in which it is embedded.  We wish to go one step further and postulate that cosmological time -- the time measured by a fundamental, comoving clock -- depends on the dynamical state of the Universe in which it is embedded.

In section 2, we discuss coordinate time, cosmological time and entropy. Section 3 considers the entropy budget of the Universe.  In section 4, we use a simple method to derive the relation between cosmological time and coordinate time and we discuss some of its consequences.  In section 5, we show that the varying flow of cosmological time does not result in an additional redshift.  Finally, in Section 6, we compare the predictions of our model to those of the $\Lambda$CDM model and test it on Type Ia Supernov\ae\  data.

\section{Cosmological time and coordinate time}

When applying GR to the whole Universe in order to derive the $\Lambda$CDM model, one assumes that the Universe is homogeneous on large scales and that the small scale inhomogeneities have no impact on the evolution of the Universe as a whole.  One also assumes that the 4-dimentional space-time can be sliced into 3-dimensional space-like hypersurfaces labeled with a time parameter $t$.  This time parameter is usually taken as the time measured by fundamental, comoving, clocks \citep{hobson2006}.  The time $t$ is then identified with the proper time of fundamental observers.  It is generally taken for granted that, if the Universe is homogeneous, this proper time flows at the same rate all along the history of the Universe. When Einstein derived the basic theory of GR, the Universe was believed to be essentially static and there was no reason to question that hypothesis.  However, in the framework of an evolving Universe, one might ask whether the rate at which time flows could not depend on the dynamical state of the Universe itself.  We thus propose to distinguish between two different times: (1) the (conventional) coordinate time parameter $t$, which is the one measured by our present clocks and is assumed to flow at a constant rate and (2) the cosmological time $\tau$, which is the one measured by fundamental clocks at any times, may depend on the state of the Universe and controls all physical processes.  When building our cosmological model, this proper time $\tau$ may not be used as an independent coordinate, as it is a function of other parameters and, thus, an emergent property. 

It is often argued that the direction of time flow, the arrow of time, is dictated by the second law of thermodynamics: the direction of time flow is the one for which the entropy of an isolated system increases.  Let us consider fundamental observers in the Universe.  The largest system possibly interacting with such observers is the region of the Universe that is causally connected to them.  That region is bounded by the particle horizon, defined by the distance a light signal could travel from the beginning of time to the observers.  From their very point of view, that system can be considered isolated since no interaction can happen with objects located further than the particle horizon.  A very simple assumption we could thus make is that the proper time $\tau$ measured by such observers is proportional to the entropy of the region of the Universe which is causally connected to them.

\section{Entropy of the Universe}

Conventional estimates of the entropy budget (e.g.\  \citealt{egan2010}) result in back holes dominating by far the entropy of the Universe.  Indeed, if one accepts the Bekenstein-Hawking estimate for the black hole entropy  \citep{bekenstein1973,hawking1976}, the entropy budget of the stellar black holes would exceed the entropy budget of the (much more numerous) stars by 17 orders of magnitude and that of the cosmic microwave background (CMB) by 7 to 8 orders of magnitude \citep{egan2010}.  Even more dramatically, a handful of supermassive black holes as those found at the centres of galaxies would completely dominate the entropy budget of the Universe.

We do not consider these possible sources of entropy in our calculations for the following reason.  Let us consider the typical situation in which a stellar black hole can be detected.  This is a binary stellar system in which one of the stars is a red giant filling its Roche lobe and loosing matter to its compact companion.  An accretion disk forms around the compact object and emits high energy radiation as its temperature reaches high values close to the central object.  Whether the compact object is a neutron star or a black hole can only be known by estimating its mass and verifying whether or not it exceeds the maximum mass of neutron stars ($\sim 3 M_{\odot}$).  The effect of a black hole just above the mass limit on its surroundings -- and, even more, on the causally connected region -- hardly differs from that of a neutron star just below the mass limit.  On the other hand, the Bekenstein-Hawking entropies would differ by many orders of magnitude without having any measurable effect on the causally connected region and, thus, on our cosmological time.  One may thus adopt one of the following reasons not to consider black holes entropy: (1) not to concur with the Bekenstein-Hawking entropy; (2) consider that the black hole entropy has no effect on the entropy budget outside of their horizons; (3) rewrite our fundamental hypothesis such that cosmological time is proportional to the entropy of radiation in the causally connected region of the Universe.

Setting black holes aside, the entropy of the Universe is dominated by the CMB \citep{egan2010}.  Relic neutrinos are predicted to contribute nearly as much as the CMB.  However, as their entropy has the same dependence on temperature and volume as the CMB photons, including them would only change the proportionality coefficient between entropy and cosmological time, and not the relation between cosmological time $\tau$ and coordinate time $t$.

\section{Evolution of the cosmological time}

The CMB is a photon gas very close to thermodynamic equilibrium, whose entropy $S$ only depends on its volume $V$ and temperature $T$ via \citep{egan2010}
\begin{equation}
   S = \frac{4 \pi^2 k_B^4}{45 c^3 \hbar^3} V T^3,
\end{equation}
where $k_B$ is the Boltzmann constant, $c$ the speed of light and $\hbar$ the reduced Planck constant. 

$V_{\rm horiz}$ being the causally connected volume, we assume $\tau \propto S(V_{\rm horiz}) \propto V_{\rm horiz} T^3$ and, as $T$ is inversely proportional to the scale factor $R$ of the expanding Universe, $\tau \propto V_{\rm horiz} R^{-3}$.  As always, the scale factor $R$ is obtained by solving the Einstein equations of GR.

By construction, GR is a local theory.  Applying it to the whole Universe results in trying to use a local theory to solve a global problem.  However, if the Universe is spatially homogeneous, applying the Einstein equations in the vicinity of any observer is equivalent to applying them anywhere else, which means this local theory may be extended to any point of space.  But, in our model, this is not true for any instant in time, as the cosmological time $\tau$, i.e. the proper time of a fundamental observer, will not flow at a constant rate.

As GR has been successfully verified locally, we make the hypothesis that the Einstein equations are valid at any location in space-time, when written as a function of the local space-time coordinates $(t, r, \theta, \phi)$.  Then, under the assumptions of homogeneity and isotropy of the Universe, the interval $ds$ is given by the Robertson-Walker metric:
\begin{equation}
   ds^2 = c^2 dt^2 - R^2 \Big[ \frac{dr^2}{1-kr^2} + r^2 (d\theta^2 + \sin^2\theta \, d\phi^2 ) \Big],
\end{equation}
where $( r, \theta, \phi)$ are the comoving spherical coordinates and $k$ is a constant measuring the spatial curvature.  The Robertson-Walker metric and the Einstein equations are thus assumed to be valid at any point in space and at any time.  However, the proper time flow of fundamental comoving observers will vary with time.  After solving the equations for any local proper time $t$, we must connect these local proper times together by replacing them by the cosmological proper time $\tau(t)$.  As already mentioned, the equations cannot be directly solved as a function of $\tau$, as the latter depends on the global properties of the Universe and is not an independent variable.  Of course, such a procedure should later be tested in a more rigorous mathematical framework, the aim of the present paper being to give a first idea of the effect of our basic hypothesis on the evolution of the Universe.

The Einstein equations in $t$ reduce to the usual Friedmann equations.  The solutions are very simple in the case of a flat Universe $(k=0)$ with no cosmological constant.  If radiation dominates the matter-energy budget, one gets $R \propto t^{1/2}$ and, if matter dominates, $R \propto t^{2/3}$.  In both cases, $V_{\rm horiz} \propto t^3$ (e.g. \citealt{weinberg2008}).  Combining these two results, we get, in a flat matter-dominated Universe, $\tau \propto t$ and cosmological time flows at a constant rate.  However, considering matter only, our Universe is not flat but open, with a negative spatial curvature depending on the density parameter $\Omega_0$ (the actual matter-energy density measured in units of the critical matter-energy density, for which the spatial curvature vanishes).  The best estimates, considering only ordinary baryonic matter, range from  $\Omega_0 \simeq 0.03$ from an inventory of the observable Universe \citep{fukugita2004,shull2012} to $\Omega_0 \simeq 0.05$ from predictions of cosmological nucleosynthesis \citep{schramm1998}.  The variation of the cosmological time rate throughout the history of the Universe is displayed on Fig.\  1, which shows that $\tau(t)$ continuously slows down in an open Universe.

The evolution of the scale factor $R$ as a function of proper time is shown in Fig.\  2.  In the $\Lambda$CDM model, the accelerated expansion in the last $\sim 6$ billion years is due to the cosmological constant $\Lambda$ \citep{suzuki2012}, which is generally interpreted as the effect of dark energy.  In our model, it is the slowing down of the cosmological proper time $\tau$ that makes the expansion appear accelerated.

At early epochs, the Universe was radiation-dominated and essentially flat \citep{weinberg2008}.  In this case, we get $\tau(t) \propto t^{3/2}$ and thus $d\tau /dt \to 0$ as $t \to 0$.  This means that the unit of cosmological proper time $\tau(t)$, as measured in constant conventional time $t$ units, becomes infinitely large at the Big Bang.  For any observer inside the Universe, while the Universe has a finite age, the Big Bang (as measured in the time $t$ units of that observer) happened infinitely far in the past.  Obviously, this may have interesting philosophical consequences.
\begin{figure}
 \label{time}
 \includegraphics[width=0.5\textwidth]{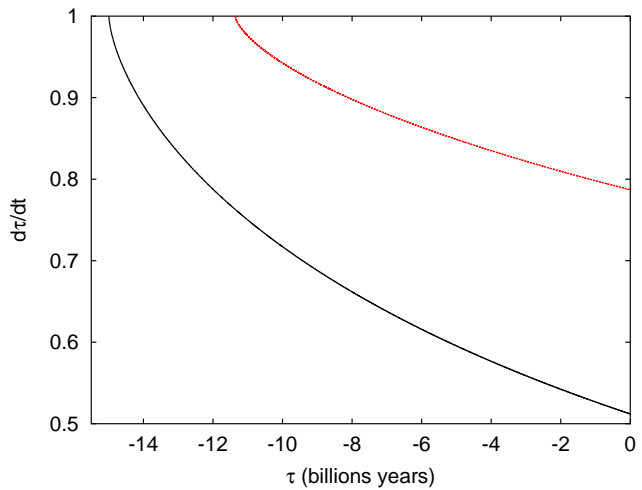}
 \caption{Variation of the cosmological time flow $d\tau /dt$ as a function of time $\tau(t)$ in an open matter-dominated Universe with different density parameters $\Omega_0$ (0.051: baryonic matter only, continuous black line; 0.32: dark matter included, dashed red line).   The cosmological time $\tau$ continuously slows down in the present matter dominated era.  The units of $\tau$ and $t$ are set equal at the beginning of the matter era, when the Universe was essentially flat and $\tau = 0$ corresponds to the present epoch.}
\end{figure}

\begin{figure}
 \label{scalefactor}
 \includegraphics[width=0.5\textwidth]{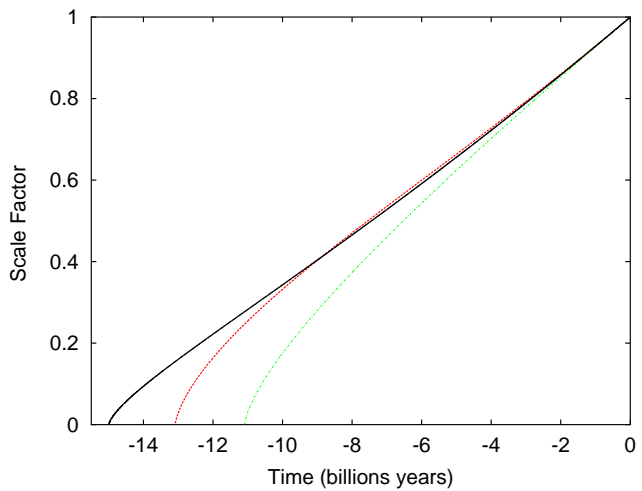}
 \caption{Evolution of the scale factor as a function of proper time in several cosmological models: the $\Lambda$CDM concordance model (dashed red curve), the same model without cosmological constant (dash-dotted green curve) and our new model (full black curve.  With $\Omega_0 = 0.051 \pm 0.014$ and a Hubble constant $H_0 = 70$ km/s/Mpc, as deduced from SNIa data (see Sect.\  6), our model predicts an age of the Universe, as measured in cosmological proper time units, of $15.0 \pm 0.5$ billion years.}
\end{figure}

\section{Cosmological redshift}

It is important to note that all physical laws have to be written as a function of the cosmological proper time $\tau$, which flows at a varying rate as the Universe evolve.  One might expect this to affect the presently observed duration or frequency of phenomena occurring far away, and thus long ago.  In fact, this is not the case, as can easily be shown, e.g., for the cosmological redshift.

For a photon emitted at cosmological time $\tau_E$ and travelling towards an observer who receives it at time $\tau_R$, $ds^2$ = 0.  Putting the source and the observer at the same $\theta$ and $\phi$ and taking into account that the time parameter $t$ has to be replaced by the proper time $\tau(t)$ at the instants of emission and reception respectively, the Robertson-Walker metric implies that
\begin{equation}
c \int_{\tau_E}^{\tau_R} \frac{d\tau}{R} = \int_{r_E}^{r_R} \frac{dr}{1-kr^2},
\end{equation}
with $r_E$ and $r_R$ being the comoving radial coordinates of the emitter and observer.  

Consider a second photon emitted at time $\tau_E + \delta \tau_E$ and received at $\tau_R + \delta \tau_R$.  It obeys a similar equation, the right hand side being equal as the emitter and observer are supposed to be at rest in the expanding space.  We thus have
\begin{displaymath}
\int_{\tau_E}^{\tau_R} \frac{d\tau}{R} = \int_{\tau_E+ \delta \tau_E}^{\tau_R+ \delta \tau_R} \frac{d\tau}{R} 
\end{displaymath}
\begin{equation}
\Rightarrow \int_{t_E}^{t_R} \frac{1}{R} \frac{d\tau}{dt}dt = \int_{t_E+ \delta t_E}^{t_R+ \delta t_R} \frac{1}{R} \frac{d\tau}{dt}dt.
\end{equation}

Assuming that the variation of $R$ is negligible during the times $\delta \tau_E$ and $\delta \tau_R$, this implies
\begin{equation}
\frac{1}{R(t_E)} \Big( \frac{d\tau}{dt} \Big)_E \delta t_E = \frac{1}{R(t_R)} \Big( \frac{d\tau}{dt} \Big)_R \delta t_R.
\end{equation}

With
\begin{equation}
\delta \tau_E = \Big( \frac{d\tau}{dt} \Big)_E \delta t_E \hspace{10pt}{\rm and}\hspace{10pt} \delta \tau_R = \Big( \frac{d\tau}{dt} \Big)_R \delta t_R,
\end{equation}
we get
\begin{equation}
\frac{\delta \tau_R}{\delta \tau_E} = \frac{\nu_E}{\nu_R} = \frac{R(\tau_R)}{R(\tau_E)},
\end{equation}
which is the usual expression for the cosmological redshift, with no additional factor due to the variation of cosmological time flow.

\section{COMPARISON WITH SUPERNOV\AE\  DATA}

The fundamental test of cosmological models is the Hubble diagram, which uses standard candles such as Type Ia Supernov\ae\  (SNIa) and compares their luminosity distances to their cosmological redshifts $z$:
\begin{equation}
z = \frac{\delta \lambda}{\lambda_0} = \frac{1}{R} - 1,
\end{equation}
where $\lambda_0$ is the rest wavelength of radiation emitted by these objects and $\delta \lambda$ its wavelength shift due to cosmological expansion.

\begin{figure}
 \label{hubble}
 \includegraphics[width=0.5\textwidth]{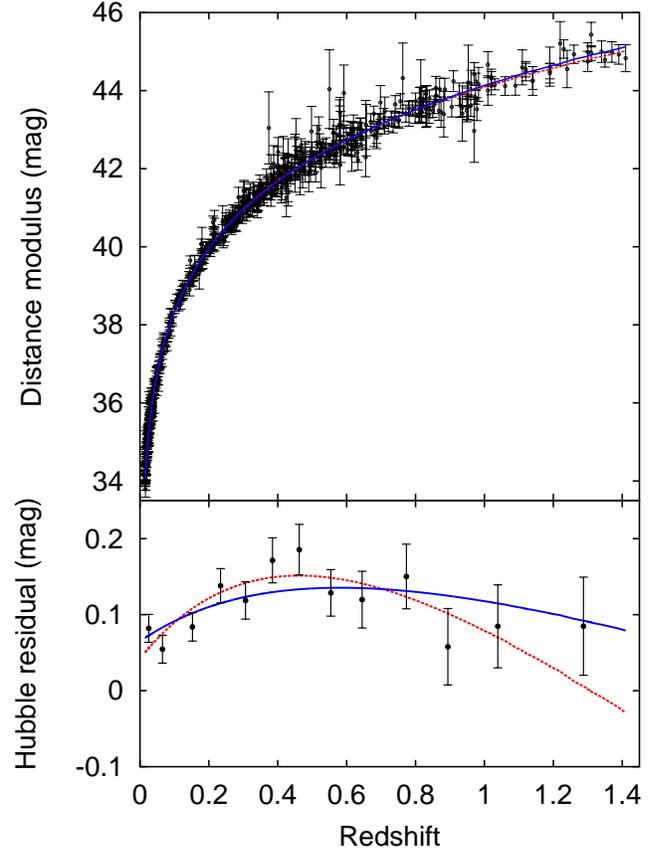}
 \caption{Comparison of cosmological models with SNIa data.  Top: Hubble diagram for 580 SNIa from the SCP Union 2.1 compilation \citep{suzuki2012} compared with the standard $\Lambda$CDM model ($\Omega_{{\rm m},0} = 0.32, \Omega_{\Lambda ,0} = 0.68$, dashed red line) and our best-fit cosmological model ($\Omega_0$ = 0.051, continuous blue line).  Bottom: same comparison after subtraction of an empty cosmological model and binning the SNIa data in bins of constant $n \delta z$, where $n$ is the number of SNIa per bin and $\delta z$ the redshift range of the bin.}
\end{figure}

To build our Hubble diagram, we consider the 580 SNIa from the Union 2.1 compilation of the Supernova Cosmology Project \citep{suzuki2012}.  These SNIa can be considered as nearly perfect standard candles once their luminosities are corrected for a so-called ‘stretch factor’ and for the SNIa color.  Indeed, their peak absolute magnitudes $M_{\rm max}$ are correlated with the slope of their post maximum brightness decrease and with their color $C$ via \citep{phillips1993,tripp1998}
 \begin{equation}
M_{\rm max} = \alpha + \beta (\Delta m_{15} - 1.05) + \gamma C,
\end{equation}
where $\Delta m_{15}$ is the decline in magnitudes during the first 15 days beyond maximum (directly related to the stretch factor; \citealt{perlmutter1997}) and $\alpha$, $\beta$ and $\gamma$ are parameters fitted on nearby SNIa \citep{tripp1998}.  Our SNIa data thus slightly differ from those of the Union 2.1 compilation, whose stretch and colour corrections use all SNIa data (not only low redshift ones) and assumes {\it a priori} that the cosmological model is a flat $\Lambda$CDM model \citep{suzuki2012}.

\begin{figure}
 \label{chi}
 \includegraphics[width=0.5\textwidth]{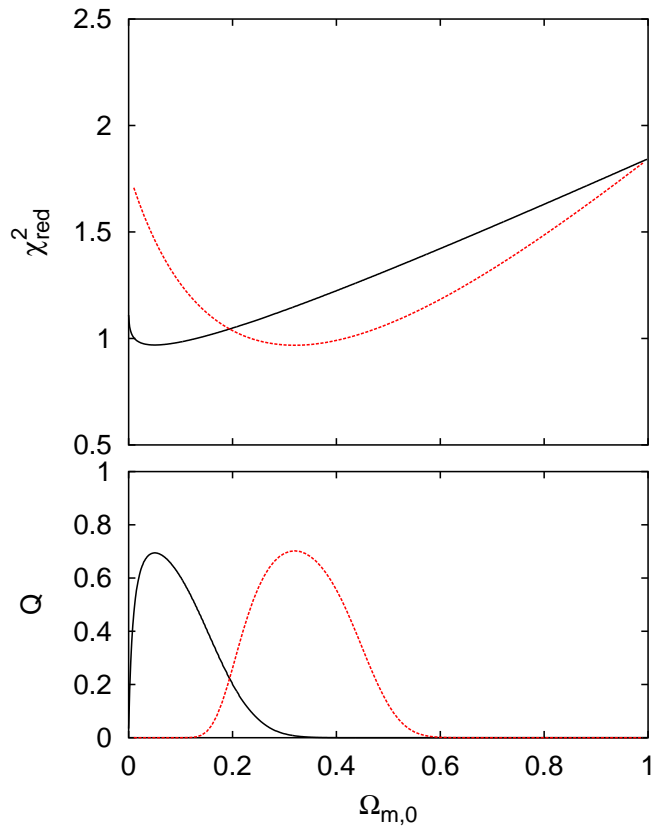}
 \caption{Quality of the fit of the cosmological models onto the SNIa data.  Top: reduced $\chi^2$ as a function of the density parameter $\Omega_0$ for the $\Lambda$CDM model (dashed red curve) and for our alternative model (full black curve).  Bottom: the corresponding probability distributions $Q$, giving the probability that a $\chi^2$ as high as the one measured is compatible with random fluctuations \citep{press1989}.  Models with low $Q$ are excluded by the data.  Our model gives  $\Omega_0 = 0.051 \pm 0.014$ (one sigma error bar), perfectly compatible with ordinary (baryonic) matter only.}
\end{figure}

The Hubble diagram for 580 SNIa \citep{suzuki2012} is displayed in Fig.\  3 together with our best fit cosmological model and the $\Lambda$CDM model.  The quality of the fits is shown on Fig.\  4, where the reduced $\chi^2$ and the associated probability distributions are plotted as a function of the present day density parameter $\Omega_0$.  It clearly shows that, while -- obviously -- the $\Lambda$CDM model requires dark matter to fit the SNIa data, our model’s best fit $\Omega_0$ amounts to $0.051 \pm 0.014$, which is perfectly compatible with baryonic matter only \citep{fukugita2004,shull2012,schramm1998}, without any need for exotic dark matter.

\section{Conclusion}

From the very simple assumption that the proper time of fundamental observers is proportional to the entropy of the region of the Universe causally connected to them, we obtain a cosmological model that perfectly fits the SNIa data without recourse to exotic dark matter nor dark energy.

Of course, more tests are needed.  In particular, the effects of the slowing down of proper time on the rotation curves of spiral galaxies and on the velocity dispersions of galaxies in clusters have to be investigated.  Qualitatively, one would predict an apparent acceleration -- similar to the one observed for the scale factor $R$ -- which might explain the larger than expected observed velocities.

Our model also leaves much more time for the formation of structures in between the time of the matter-radiation decoupling $(z \sim 1000)$ and the earliest galaxies observed $(z \sim 10)$.  Whether or not some dark matter would still be needed to accelerate the baryonic matter condensation has also to be investigated.

Cosmological nucleosynthesis and CMB anisotropies also have to be studied in the framework of the present model, for it to become a viable alternative to the $\Lambda$CDM model.

\bsp

\label{lastpage}


\begin{thebibliography}{99}
\bibitem[\protect\citeauthoryear{Bekenstein}{1973}]{bekenstein1973}Bekenstein J.D., 1973, Phys. Rev. D 7, 2333
\bibitem[\protect\citeauthoryear{Egan \& Lineweaver}{2010}]{egan2010}Egan C.A. \& Lineweaver C.H.,  2010, ApJ, 710, 1825
\bibitem[\protect\citeauthoryear{Einstein}{1916}]{einstein1916}Einstein A., 1916, Ann. Phys. (Berlin), 354, 769
\bibitem[\protect\citeauthoryear{Fukugita \& Peebles}{2004}]{fukugita2004}Fukugita M. \& Peebles P.J.E., 2006, ApJ, 616, 643
\bibitem[\protect\citeauthoryear{Hawking}{1976}]{hawking1976}Hawking S.W., 1976, Phys.\  Rev.\  D 13, 191
\bibitem[\protect\citeauthoryear{Hobson et al.}{2006}]{hobson2006}Hobson M.P., Efstathiou G.P., Lasenby A.N., 2006,  General Relativity, Cambridge University Press, New York
\bibitem[\protect\citeauthoryear{Milgrom}{1983}]{milgrom1983}Milgrom M., 1983, ApJ, 270, 365
\bibitem[\protect\citeauthoryear{Perlmutter et al.}{1997}]{perlmutter1997}Perlmutter S.\  et al., 1997, ApJ, 483, 565
\bibitem[\protect\citeauthoryear{Phillips}{1993}]{phillips1993}Phillips M.M., 1993, ApJ, 413, L105
\bibitem[\protect\citeauthoryear{Press et al.}{1989}]{press1989}Press W.H., Teukolsky S.A., Vetterlink W.T. \& Flannery B.P., 1989, Numerical Recipes, Cambridge University Press, New York
\bibitem[\protect\citeauthoryear{Schramm \& Turner}{1998}]{schramm1998}Schramm D.N. \& Turner M.S., 1998, Rev.\  Mod.\  Phys., 70, 303
\bibitem[\protect\citeauthoryear{Shull et al.}{2012}]{shull2012}Shull J.M., Smith B.D. \& Danforth C.W., 2012, Ap,J 759, 15
\bibitem[\protect\citeauthoryear{Suzuki et al.}{2012}]{suzuki2012}Suzuki N. et al., 2012, ApJ, 746, 24
\bibitem[\protect\citeauthoryear{Tripp}{1998}]{tripp1998}Tripp R., 1998, A\&A, 331, 815
\bibitem[\protect\citeauthoryear{Weinberg}{2008}]{weinberg2008}Weinberg S., 2008, Cosmology, Oxford University Press, New York

\end{thebibliography}
\end{document}